# Opals as Colorful Radiative Coolers

*Hyeon Ho Kim*[1], *Eunji Im*[2], and *Seungwoo Lee*[1,2,3*]

[1]KU-KIST Graduate School of Converging Science and Technology, Korea University, Seoul 02841, Republic of Korea
[2]Department of Biomicrosystem Technology, Korea University, Seoul 02841, Republic of Korea
[3]KU Photonics, Korea University, Seoul 02841, Republic of Korea

*Email: seungwoo@korea.ac.kr

**Abstract**: Radiative cooling has proven to be a powerful strategy for sustainable thermal management. Nanophotonic structures enabling broadband reflection lead to minimization of sunlight absorption, which has brought nighttime-limited radiative cooling into daytime applications. However, this broadband reflection strategy in turn restricts the accessible colorization of radiative coolers to white or neutral, consequently hindering their practical applications, particularly for aesthetic purposes. With a few exceptions, selective absorption at a specific visible wavelength has been the most prevalent paradigm for colorization of radiative coolers. However, this absorption-based colorization inevitably make the radiative cooler prone to heating, thus decreasing the cooling efficiency. Here, we demonstrate an undiscovered usage of opals for advancing color-preserved daytime radiative coolers. Opals, which have served mainly as Bragg reflective color pigments thus far, can be considered an effective homogeneous medium in the mid-infrared region. Thus, opals can also be envisioned as reflectively colorful metamaterials capable of radiative cooling even under the direct summer sun. Together with the soft fluidity of colloidal suspensions, opals can serve as platforms for easy-to-craft, large-scale, and colorful radiative coolers with minimal solar absorption.

ToC:

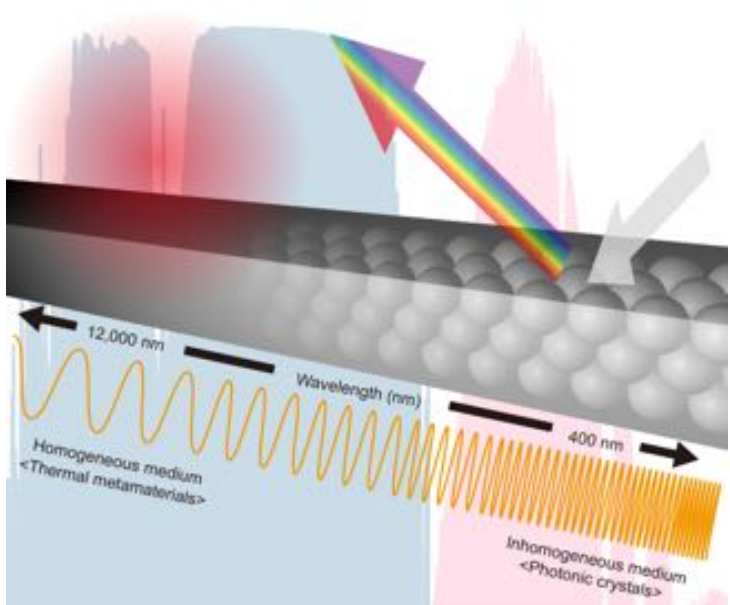

Introduction: Over the past few years, daytime radiative cooling has attracted radically increased interest from both the nanophotonics and energy societies following its first realization in 2014 by S. Fan et al.[1,2] Given the intrinsic phononic vibration and emissivity of the specific materials at mid-infrared (IR) wavelengths light (i.e., 8 ~ 13 μm wavelengths, where the atmosphere is almost transparent),[1,3-6] the ability of rationally designed nanophotonic structures to efficiently reflect the full solar spectrum has enabled radiative cooling even in the daytime without any consumption of energy.[7-23] Since then, nanophotonics has been considered a versatile and deterministic strategy not only for expanding the range of light-matter interactions but also for reducing the energy demand for cooling.

Much of the development of daytime radiative cooling has been oriented towards addressing two main challenges: (i) improving the cooling efficiency by optimizing the device and nanophotonic architectures[1,2,4,7,9,19] and (ii) translating daytime radiative cooling into practical applications (e.g., a building material) by expanding the available material libraries.[5,6,11,12,17,18,21,22] In particular, a variety of unconventional soft nanophotonic materials, such as paint,[11,17] bead-polymer composites,[5] textiles,[12,18] and celluloses,[22] all of which could be processed in a scalable and cost-effective manner, were found to simultaneously scatter incoming solar energy and emit mid-IR light through passive radiation. These recent findings could facilitate practical and scalable realization of daytime radiative coolers.

Nevertheless, colorization of daytime radiative coolers while maintaining high cooling efficiency remains an obstacle to their practical application. The aesthetic colorization could meet a pivotal technological demand for the rich application space of daytime radiative cooling. However, efficient daytime radiative cooling and colorization appear to be mutually exclusive because preservation of the colors has been generally achieved by absorption of a specific visible wavelength, which results in heating of the structures.[7,16,20,24] Indeed, most daytime radiative coolers reported thus far have a white or neutral color, because mainly focusing on improving the radiative cooling efficiency.[2,4,6,12,17,18,22] A few exceptions have been developed in conjunction with theoretical upper and lower bounds of color-preserved daytime radiative coolers;[7,16,20,24] however, these studies are still limited to absorption-based colorizations.

Here, we introduce that reflective full colorization (i.e., red, green, and blue) with minimized solar absorption can be massively and economically translated into practical daytime radiative coolers by benefiting from soft self-assembly. As a case study, we implement self-assembled silica opals on a silicon (Si) wafer to effectively address the above paradoxical paradigm. The design principles outlined by both wavelength-scale diffractive optics and deep-subwavelength-scale homogenization theory can generalize the reflective colorization of a radiative cooler, making this colorization attractive not only for expanding the fundamental limit of color-preserved radiative cooling but also for diversifying the materials and processing platforms.

Results and Discussion: **Figure 1a** outlines the principle underpinning how opals can act as a color-preserved radiative material with high cooling efficiency. Herein, we use opals assembled from 300 nm silica colloidal nanospheres (NSs) as a representative example (opals in an air medium). Opals are generally referred to as close-packed face-centered-cubic (FCC) crystals with 74 vol%.[25-31] In the visible spectrum (400 nm ~ 760 nm wavelengths), the primitive cell of opals, comparable to the wavelength of interest, can be viewed as a photonic crystal (i.e., an inhomogeneous medium), exhibiting iridescent colors resulting from Bragg diffractions.[27,28,31] In other words, preservation of colors can be achieved via structural diffraction rather than selective absorption. Meanwhile, opals become a deep-subwavelength-scale material in the mid-IR region; consequently, they are transformed into a homogeneous medium (known as metamaterials), which enables radiative thermal loading and emission. These two aspects of opals are detailed as follows.

In the visible region, the bandgap of opals can be opened by (111) Bragg diffraction (from the Γ to L points), as shown in the numerically calculated band diagram (**Figure 1b**); the position of this bandgap can be precisely tuned over the entire visible range merely by adjusting the size of the silica NSs (from 200 to 300 nm). In the mid-IR region, the effective parameters of silica opals can be numerically retrieved using homogenization theory (**Figure 1c**).[32-35] To this end, we numerically retrieve the real ($n$) and imaginary ($k$) refractive indices by using the *s*-parameters, as detailed in **Figure S1**, Supporting Information. **Figure 1c** summarizes the obtained effective $n$ and $k$ of silica opals (e.g., 300 nm silica NSs) at the 6 ~ 14 μm wavelengths. Notably, significant absorption ($k$) is clearly visible between the 8 and 14 μm wavelengths, coinciding with the Kramers–Kronig relation.[36,37] Additionally, regardless of the silica NS size, consistent $n$ and $k$ of opaline films are observed (see **Figure S1**, Supporting Information). These theoretical results support that silica opals can actually be a homogenous medium in the mid-IR region that can radiatively load and emit thermal energy. Note that both the effective $n$ and $k$ of silica opals in the air host medium are lower than those of the fused silica solid film due to their lower vol%.

Give the above nanophotonic and thermal analyses of silica opals, we numerically calculated the absorption spectra of 150 μm thick silica opals, as summarized in **Figure 1d**. The finite-difference time-domain (FDTD) method was used to this end (see details in the Supporting Information). Note that absorptivity is equal to emissivity, according to Kirchhoff's law. As expected, a strong absorption is theoretically predicted at the 8 ~ 13 μm wavelengths, evidencing that silica opals can act as radiative coolers. Notably, silica opals negligibly absorb sunlight, whereas colorization is still available via Bragg diffraction, as already probed above. Therefore, this diffractive colorization of opals can minimize the heating effect, possibly caused by absorption of sunlight, which makes them starkly different from the previously reported, color-preserved radiative coolers.[16,20,24]

Towards the realization of radiative coolers, we self-assemble silica opals by drop-casting (see more details in **Figure S2**, Supporting Information). Highly concentrated suspensions of silica colloidal particles with diameters of 200 ~ 290 nm are dropped onto the entire area of the substrate (3.0 cm by 3.0 cm) and slowly dried. Then, close-packed silica FCC crystals are obtained. To maximize the radiative cooling performance, relatively thick opals are assembled (i.e., 150 μm). Herein, we use a double-side-polished crystalline Si (*c*-Si) wafer with a 550 μm thickness as a representative cooling target because it has been widely used as a foundational material for outdoor optoelectronics.[38] However, under sunny outdoor conditions, *c*-Si and the relevant devices are generally heated up to 50 °C or even higher,[6,39] giving rise to performance degradation.[39] Therefore, *c*-Si outdoor optoelectronics could exhibited improved performance and aesthetic colorization through the use of such versatile coatings of opals. A more detailed perspective on opaline *c*-Si optoelectronics will be reported separately, which is beyond the main scope of this work. A thin polydimethylsiloxane (PDMS) polymer layer with a thickness of less than 5 μm is used as an adhesion promoter between the silica opals and *c*-Si substrate. The silica opals without the PDMS layer are readily delaminated from the *c*-Si substrate.

**Figures 2a-e** show macroscopic and microscopic images of reddish (R), greenish (G), and bluish (B) silica opals on PDMS/*c*-Si stacks, assembled from 200 nm, 240 nm, and 290 nm NSs (see the scanning electron microscopy (SEM) images, presented in **Figures 2c-e**). The neutrally colored *c*-Si is transformed into a distinct color (**Figure 2a**). The cross-sectional optical microscopy (OM) image (**Figure 2b**) confirms the 150 μm thickness of the assembled opals (e.g., 200 nm NSs). The corresponding normal reflection spectra ((111) direction), presented in **Figure 2f**, further support that such colorization originates from diffractive reflection, matching well with the theoretically predicted bandgap (see **Figure 1b**).

The absorptivity of RGB silica opals in the visible and NIR regions, confirmed by UV/Vis absorption (Cary 5000, Agilent Technologies) and Fourier transform (FT) IR spectroscopic (Cary 630, Agilent Technologies) analysis with an integrating sphere, is not negligible, in contrast to the theoretical expectations (**Figure 3a**). We attribute this nonnegligible broadband absorption to light localization (i.e., Anderson localization) caused by the imperfections of the self-assembled opals, such as vacancy defects, line cracks, and amorphous areas (**Figure S3**, Supporting Information).[40,41] This absorption is minimized at the photonic bandgap position, as shown by the dips in the absorptivity spectrum (**Figure 3a**). Despite the nonnegligible absorptivity, colorization of the silica opals is achieved with much smaller solar absorption than that of the previously reported radiative coolers.[16,20,24]

More critically, these silica opals can absorb a considerable amount of mid-IR light, agreeing well with the numerical predictions. Figure 3b shows the representative results of mid-IR absorption; regardless of the Bragg diffractive color, similar absorptivity over the mid-IR region is observed because such RGB opals with 200 ~ 290 nm silica NSs can be almost considered as an effective medium (as already confirmed in **Figure 1c** and **Figure S1**, Supporting Information). Indeed, silica opals were found to cool the substrate (i.e., plastic petri-dish made of crystal-grade polystyrene with negligible mid-IR absorptivity) below the ambient temperature in dark indoor room (**Figure 3c**). Taken together, these results show that our silica opals indeed play dual roles as a radiative cooler, with (i) reflective color pigments in the visible region (i.e., an inhomogeneous medium as photonic crystals) and (ii) the radiative thermal loader and emitter in the mid-IR region (i.e., a homogeneous medium as thermal metamaterials).

Finally, we profile the outer daytime radiative cooling performance of our devices using a location in Seongbuk-gu, Seoul, as a case study (i.e., on the ground in front of the KU R&D center on the Korea University campus at 37º35'24.0''N 127º01'36.4''E, as shown in **Figure 4a**). As with the previously reported standards, the samples are encapsulated by thermally reflective metalized film (i.e., aluminized polyethylene terephthalate (PET)) except for a low-density polyethylene (LDPE) window (**Figure 4b**) and exposed to both sunny and cloudy conditions (see more details in **Figure S4**, Supporting Information). Even under direct sunlight, we can see distinct structural colors (**Figure 4c**).

**Figure 5a** shows representative results of the daytime-traced temperature measurements across the samples, including (i) *p*-doped *c*-Si (black line), (ii) PDMS/*p*-doped *c*-Si (mellow line), and (iii) differently colored silica opals/PDMS/*p*-doped *c*-Si. These measurements were performed on August 5$^{th}$, 2019 from 10:00 to 16:00 as a case study of direct summer sun (i.e., a sunny day with temporary cloudy conditions). The corresponding solar irradiance is indicated by green lines. Several features are noteworthy as follows.

First, since the 550 μm thick *c*-Si can fully absorb sunlight, its temperature becomes higher than 60 ºC. The coating of the thin PDMS layer negligibly influence on the radiative cooling of *c*-Si. Even if the siloxane bond of PDMS can broadly absorb mid-IR light via molecular vibrations,[6] we controlled the thickness of the PDMS to be as thin as possible by spin-coating it to minimize the role of the PDMS in daytime radiative cooling (~ 5 μm).

Second, and more importantly, the opals significantly reduce the temperature of the *c*-Si by as much as 13 ºC (e.g., reddish opals), for example, at approximately 11:00 AM (**Figure 5b**). This result implies that radiative thermal loading of silica opals, already confirmed in **Figure 3b**, can indeed cool the structures in a radiative fashion. Additionally, the diffusive localization of incoming sunlight within the assembled opals can further contribute to daytime radiative cooling because less solar power should infuse into the *c*-Si.

Third, the daytime radiative cooling due to the reddish opals outperforms that due to the greenish and bluish opals. According to the solar spectrum shown in the pink map of **Figure 1d**, the amount of incoming sunlight is reduced in the order of bluish, greenish, and reddish. Therefore, sunlight can be more reflected by reddish opals compared with greenish and bluish opals. This dependency of the cooling performance on the reflective colors is mitigated under indirect sunlight (e.g., temporary cloudy conditions (see **Figures S6-S7**, Supporting Information), which in turn further evidences that the higher daytime cooling performance of the reddish opals originates from more efficient solar reflection compared with the greenish/bluish counterparts.

Fourth, the radiation effect becomes clearer, as the temperature of the *c*-Si increases. The corresponding radiative power, experimentally obtained during this daytime-traced, temperature measurement, is summarized in **Figure 5c** (see details of the calculation in the Supporting Information). This result occurs because blackbody radiation can be facilitated with an increase in temperature. Accordingly, the net cooling power could be higher at higher device temperatures and lower thermal coefficients, as shown in **Figure 5d** and **Figure S5**, Supporting Information. Note that such cooling characteristics of opals are reproducibly observed under different wheather conditions (e.g., cloudy day), as summarized in Supporting Information (**Figures S6-S7**).

Conclusions: In conclusion, we show for the first time that silica opals can simultaneously satisfy the favorable *colorization* and *thermal loading* for daytime radiative cooling. Their high compatibility with large-scale and massive processing, benefiting from the soft fluidity of the colloidal suspension, provides an additional advantage in terms of their practical use in aesthetic applications. The versatile drop-casting and successive drying allow us to efficiently coat thick silica opals on a solar absorbing substrate over the centimeter scale (at least). Surprisingly, daytime radiative cooling by as much as 15 ºC, while maintaining the nonabsorbing colorization is achieved through this versatile coating of the colloidal suspension. Using this structural colorization design strategy as a starting point, many other different, color-preserved radiative coolers could soon be developed, because the visible and thermally emissive photonic crystal motif has been well diversified over the past two decades. Then, the real-world applications of daytime radiative coolers could be unprecedentedly expanded along with significant advances in such versatile realization.

Experimental sections:

Synthesis of silica colloids. All chemical reagents, including Tetraethyl orthosilicate (Aldrich, ≥99.0%), ammonia solution (Aldrich, 25%), and anhydrous ethanol (Aldrich, ≥99.5%) were used as received. Deionized (DI) water (Milli-Q water) was used as a solvent for the synthesis of silica nanospheres. We synthesized silica nanospheres in a one-step process using the solvent varying method, which is modified the sol-gel Stöber method. Given the mixture of TEOS, ammonia solution, and DI water (respectively 6, 8, and 3 mL) we varied the volume of ethanol from 47 to 68 mL. Through this, the size of silica nanospheres were precisely tuned from 200 nm to 290 nm. The reaction temperature was 60 °C and the mixture was stirred at a speed of 800 rpm using a magnetic bar for 2 hours. The synthesized silica nanospheres was spun at 300 rcf for 30 min in a centrifuge and washed with ethanol 5 times. After the ethanol was completely evaporated, the silica particles were dissolved in pure ethanol with a concentration of 10 wt%.

Assembly of opals. We assembled opals using gravity sedimentation through solvent evaporation (so called drop-casting). A 3 cm x 3 cm crystalline-silicon (*c*-Si) substrate, which was already washed with isopropyl alcohol (IPA), acetone, ethanol, and DI water in successive way, was coated with polydimethylsiloxane (PDMS) using spin coating (9000 rpm for 60 s) and thermally cured in 80°C oven for 1 h. Then, 1 mL of a 10 wt% silica colloidal solution was drop-casted on the PDMS/*c*-Si substrate in a 24°C oven. During the evaporation of ethanol, silica nanospheres were self-assembled into opaline structures (see **Fig. S2**).

Experiments on radiative cooling. The measurement of outdoor radiative cooling was carried out in Seongbuk-gu, Seoul with 37°35'24.0"N 127°01'36.4"E (Korea University campus). Resistance temperature detectors (SA-1 RTD, Omega Korea) were attached to the backside of the samples to measure the temperature and connected to a data logger (OM-CP-OCTRCT, Omega Korea) to record the time-traced temperature variations. The RTD-attached samples were put into the custom-built, thermally isolated box (**Fig. S4**), consisting of paper box, aluminized polyethylene terephthalate (PET) film, styrofoam, acrylic box, and low-density polyethylene (LDPE) film (see **Figure 4b** of main manuscript). As a windshield, the LDPE seals the window. A pyranometer (TES1333R) was placed towards the sky to measure the solar irradiance incident on the sample

Acknowledgement

This work was supported by Samsung Research Funding & Incubation Center for Future Technology of Samsung Electronics (Project Number SRFC-MA1801-04). KU-KIST school project supported graduate student scholarship for H.H.K. The authors also thank Ji-Hyeok Huh and Kwangjin Kim for their helpful assistance of spectroscopic characterizations and effective medium theoretical analysis.

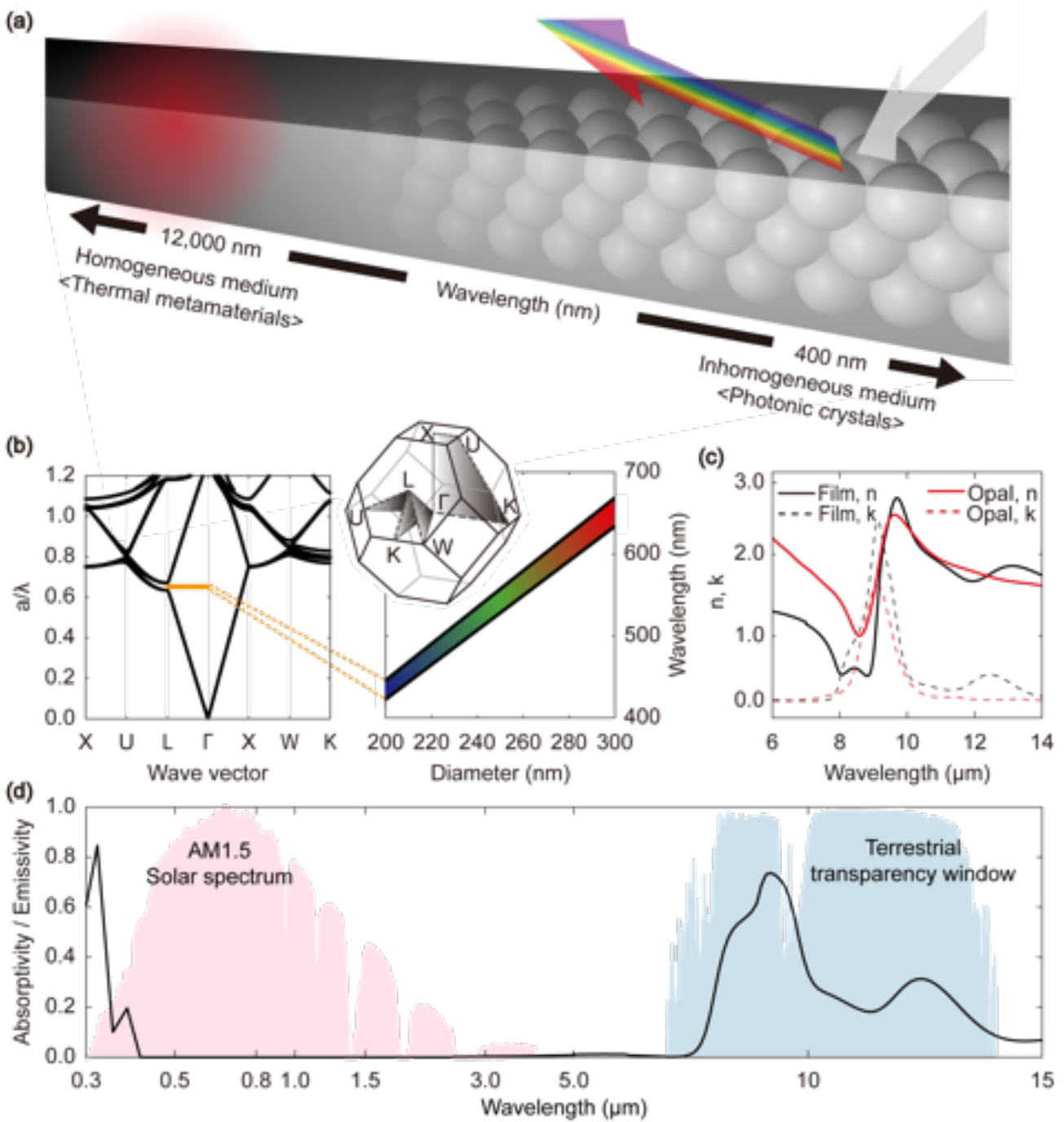

**Figure 1.** (a) Design principle for colorful radiative coolers of opaline crystals. In the visible region, opals can be viewed as an inhomogeneous medium, as the length scale of their unit cell is comparable to the wavelength of visible light. Therefore, Bragg diffractive colorization can occur via the inhomogeneous medium effect (i.e., photonic crystal effect). In contrast, such a structural length scale becomes negligible in the mid-infrared (IR) region, where opals can be considered as a homogeneous medium and thus a thermal metamaterial. (b) Photonic bandgap diagram of opals, numerically predicted by the finite-difference time-domain (FDTD) method. According to the diameter of the silica nanospheres (NSs) to be assembled into opals, the photonic bandgap along the Γ to L points (i.e., (111) diffraction) can be widely tuned from reddish to bluish colors. (c) Effective refractive indicies of silica opals (e.g., assembled by 300 nm silica NSs) in the mid-IR region, which are theoretically predicted by the s-parameter retrieval method.[32-35] (d) Theoretically predicted absorptivity/emissivity of 150 μm thick silica opals from the UV to mid-IR regions.

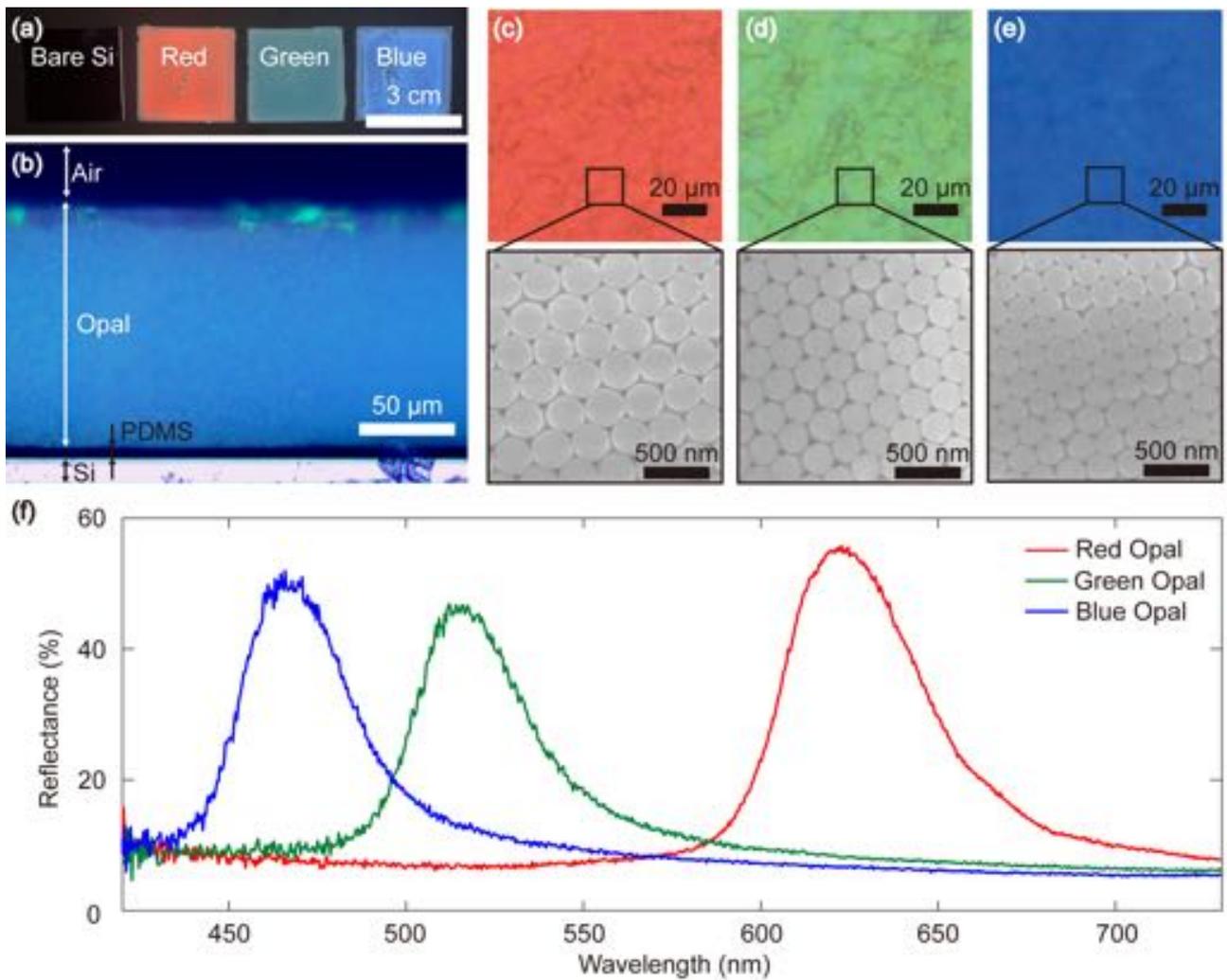

**Figure 2.** Bragg reflective colorization of opals, self-assembled on a polydimethylsiloxane (PDMS)/*p*-doped crystalline silicon (*c*-Si) wafer (substrate). (a) Macroscopic digital camera image of a bare substrate and reddish/greenish/bluish opals on a substrate (from left to right). (b) Cross-sectional optical microscopy (OM) image of bluish opals (bright field mode). The opals used in this study are 150 μm thick. (c-e) Bright field OM (top panel) and scanning electron microscopy (SEM) images of (c) reddish, (d) greenish, and (e) bluish opals, self-assembled from 290 nm, 240 nm, and 200 nm silica NSs, respectively. (f) Corresponding reflection spectrum of opals taken along the normal direction (coinciding with the direction from the Γ to L points of the photonic band diagram).

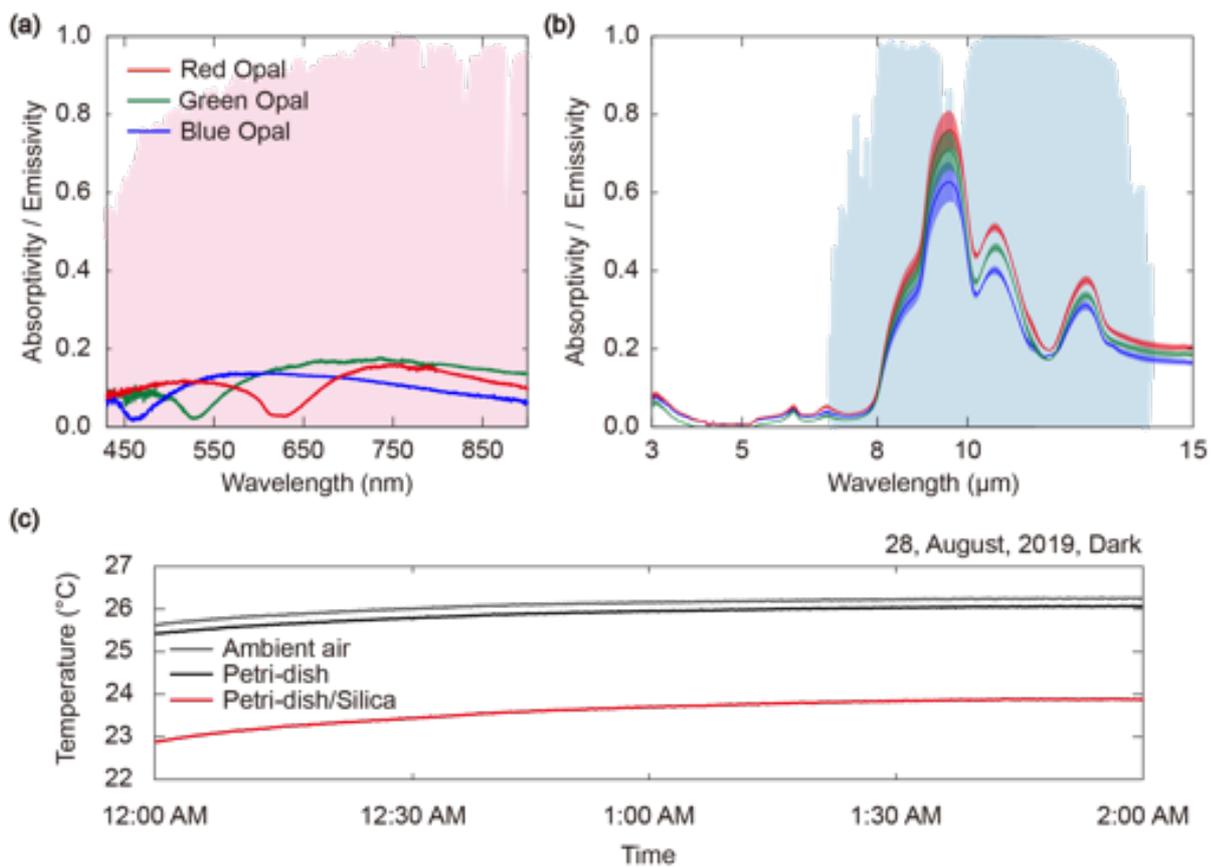

**Figure 3**. (a-b) Experimentally measured absorptivity/emissivity of bluish, greenish, and reddish opals in the (a) visible and (b) mid-IR regions. (c) Time-trace temperature of ambient, bare substrate (plastic petri-dish composed of crystal-grade polystyrene), and silica opals/substrate stacks, which were measured in dark indoor room.

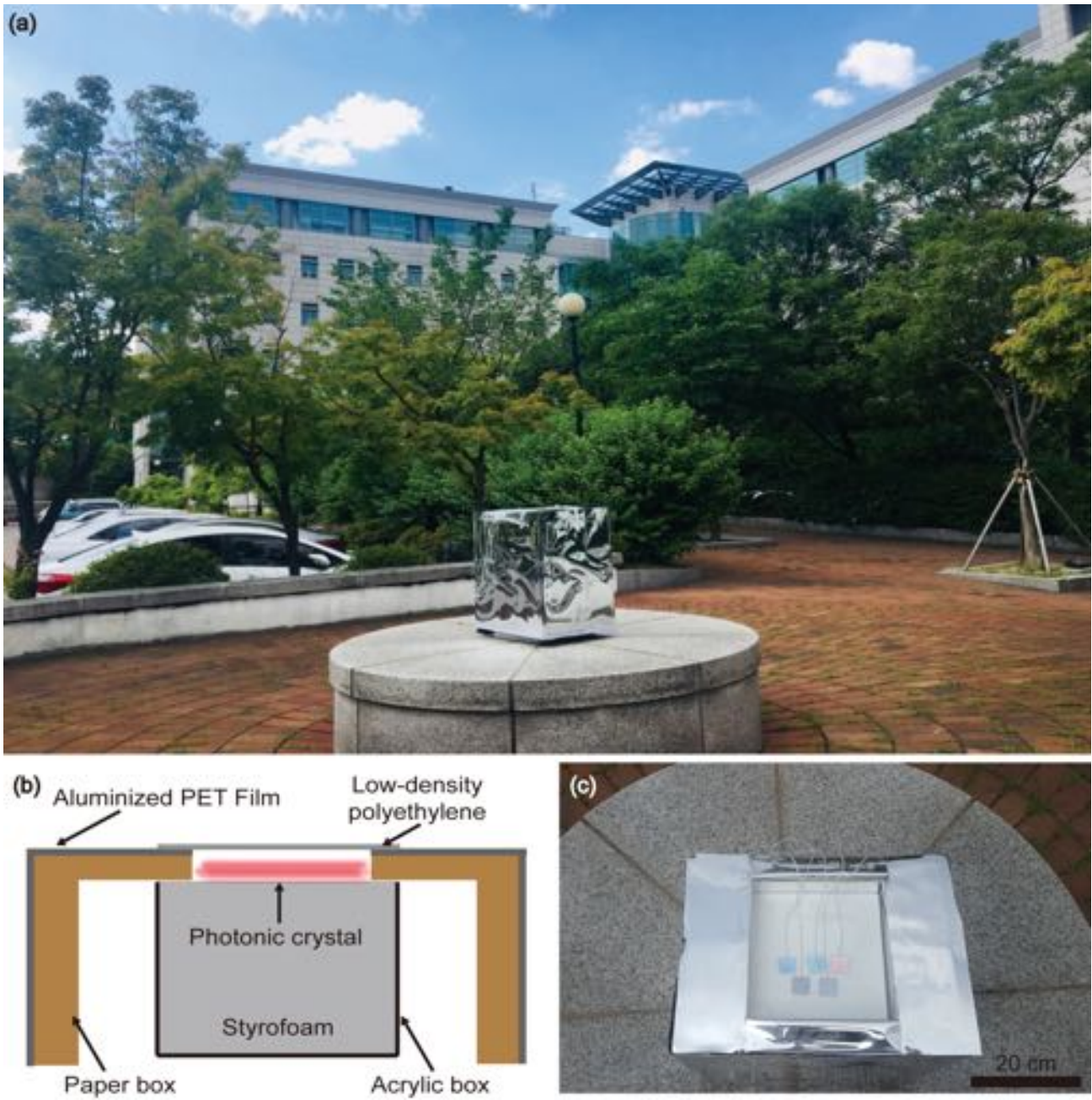

**Figure 4.** Outdoor measurement of daytime radiative cooling of assembled opals. (a) Measurement location (the front of the KU R&D center on the Korea University campus at 37º35'24.0''N 127º01'36.4''E). (b) Schematic and (c) macroscopic image of the measurement equipment. Reddish, greenish, and bluish opals on a substrate are located at the right, middle, and left positions of the top row, respectively, while *c*-Si and PDMS-coated *c*-Si are located at the left and right positions of the bottom panel, respectively. The measurement box is encapsulated by thermally reflective metalized film (i.e., aluminized polyethylene terephthalate (PET)) except for the low-density polyethylene (LDPE) window.

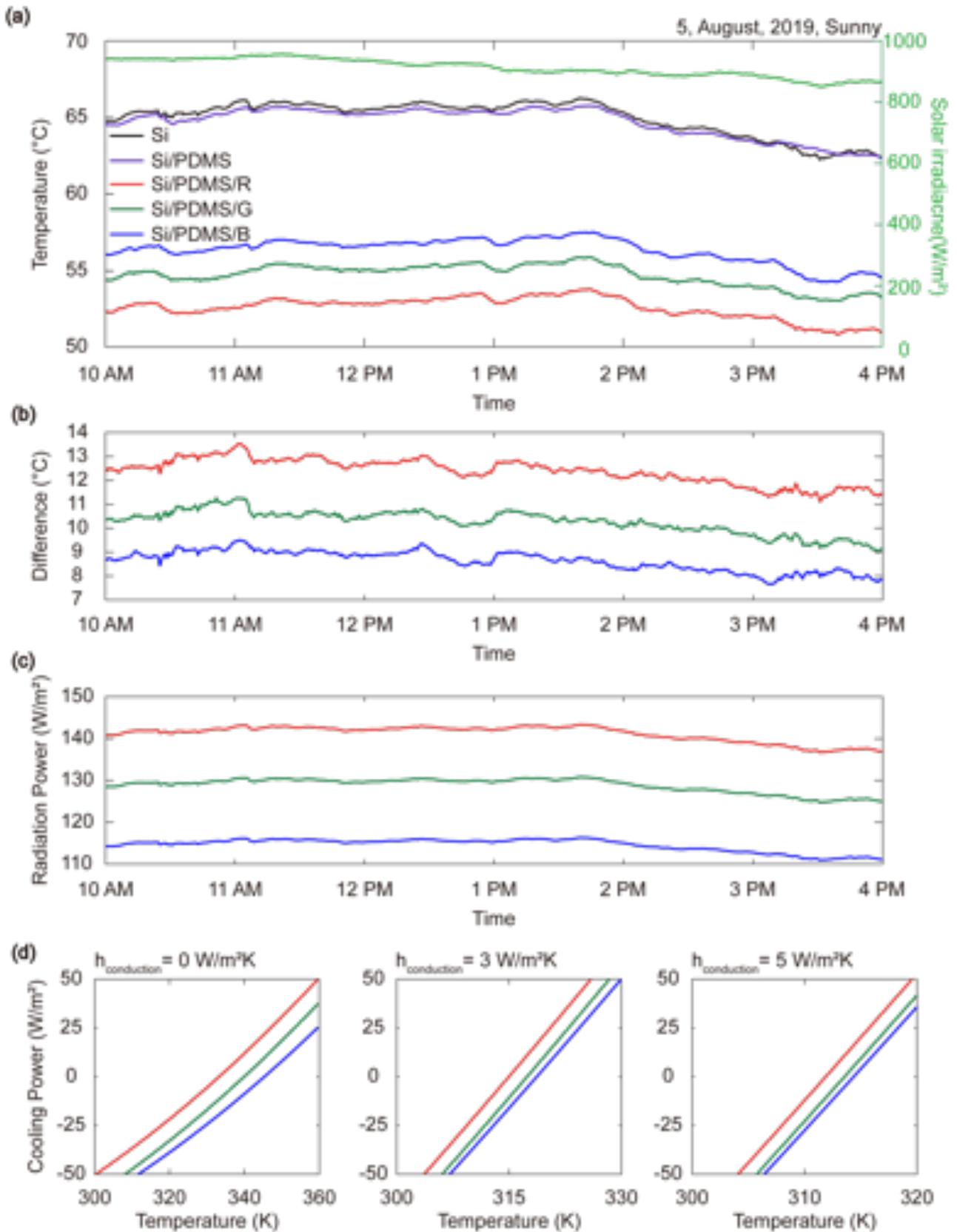

**Figure 5.** Outdoor daytime radiative cooling performance of opals measured on August 5[th], 2019 as a representative case study under direct summer sun. This day was quite sunny, but clouds temporarily passed over the sky. (a) Time-trace temperature variation. (b) Time-trace temperature difference between bare substrate and opals/substrate stacks. (c) Corresponding radiation power. (d)

Calculated net cooling power of RGB opals as a function of the device temperature and thermal coefficients.

*Supporting Information for*

# Opals as Colorful Radiative Coolers

*Hyeon Ho Kim*[1], *Eunji Im*[2], and *Seungwoo Lee*[1,2,3*]

[1]KU-KIST Graduate School of Converging Science and Technology, Korea University, Seoul 02841, Republic of Korea
[2]Department of Biomicrosystem Technology, Korea University, Seoul 02841, Republic of Korea
[3]KU Photonics, Korea University, Seoul 02841, Republic of Korea

*Email: seungwoo@korea.ac.kr
Keywords: Colloids, Opals, Radiative cooling, Photonic Bandgaps

Table of Contents:

1. Retrieval of effective refractive index by using *s*-parameters

2. Numerical calculation of absorptivity/emissivity of opals

3. Self-assembly of opals

4. Microscopic view of the cracks in self-assembled opals

5. Experiments on daytime radiative cooling

6. Calculation of radiative cooling power

7. Additional data for daytime radiative cooling of opals

1. Retrieval of effective refractive index by using *s*-parameters

Finite-difference, time-domain (FDTD) numerical simulation, supported by 2003–2015 Lumerical Solution was carried out for the theoretical analysis of the effective refractive index of silica opals at the mid-infrared (IR). In particular, we retrieved effective refractive index by using *s*-parameters.[1] The unit cell of opals is shown in **Fig. S1a-b**. Wavevector (k) of incident light was normal to (111) plane for *s*-parameter calculation. The E-field monitoring in parallel to the incident wavevector was used to set the effective thickness. Because the retrieved effective parameter is intrinsic property, the thickness of the homogeneous slab should be independent on the retrieved parameter. The retrieved effective indices for 200 nm, 240 nm, and 290 nm of silica colloidal opals are shown in **Fig. S1c**.

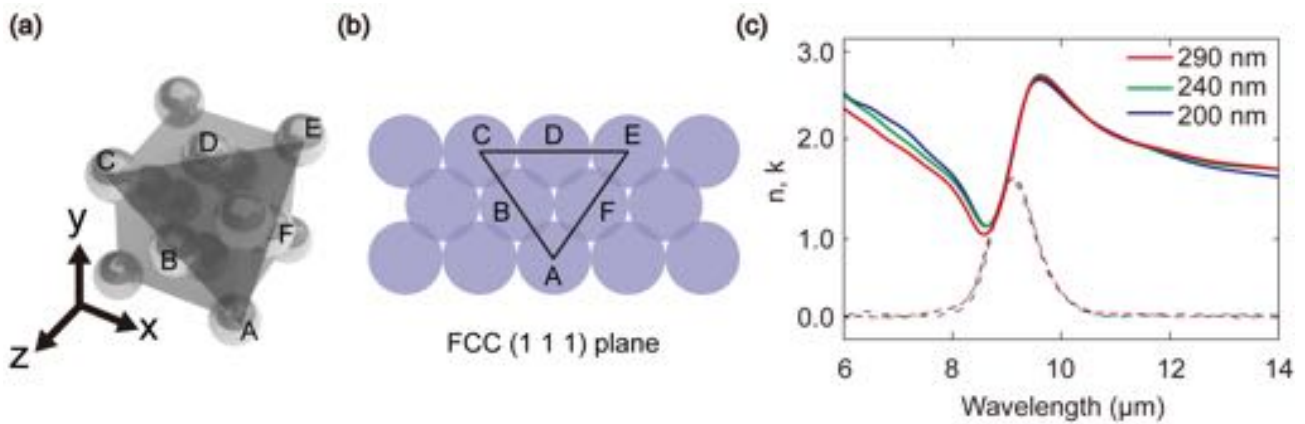

**Figure S1.** (a) Unit cell of opal with highlight of (111) plane. (b) Schematic for a front view of (111) plane. (c) Mid-infrared (IR) effective refractive index of silica opal composed of differently sized silica nanospheres (200 nm, 240 nm, and 290 nm). Solid and dotted lines indicate real ($n$) and imaginary ($k$) refractive indices respectively.

2. Numerical calculation of absorptivity/emissivity of opals

Given effective parameters obtained by the *s*-parameter retrieval method, absorptivity/emissivity of the 150 μm thick silica opals were numerically calculated by FDTD method at the mid-IR regime (3 μm to 15 μm). In this regime, opals were approximated by the thin solid film with the effective parameters, presented in **Fig. S1** and **Fig. 1c**. This can be justified by a homogenization theory under the deep-subwavelength scale of opals. From UV to NIR (3 μm wavelength), whole structure of FCC along the same thickness was modeled in this numerical simulation and the absorptivity was numerically obtained.

## 3. Self-assembly of opals

### 3-1. Synthesis of silica colloids

All chemical reagents, including Tetraethyl orthosilicate (Aldrich, ≥99.0%), ammonia solution (Aldrich, 25%), and anhydrous ethanol (Aldrich, ≥99.5%) were used as received. Deionized (DI) water (Milli-Q water) was used as a solvent for the synthesis of silica nanospheres. We synthesized silica nanospheres in a one-step process using the solvent varying method,[2] which is modified the sol-gel Stöber method.[3] Given the mixture of TEOS, ammonia solution, and DI water (respectively 6, 8, and 3 mL) we varied the volume of ethanol from 47 to 68 mL. Through this, the size of silica nanospheres were precisely tuned from 200 nm to 290 nm. The reaction temperature was 60 ° C and the mixture was stirred at a speed of 800 rpm using a magnetic bar for 2 hours. The synthesized silica nanospheres was spun at 300 rcf for 30 min in a centrifuge and washed with ethanol 5 times. After the ethanol was completely evaporated, the silica particles were dissolved in pure ethanol with a concentration of 10 wt%.

### 3-2. Assembly of opals

We assembled opals using gravity sedimentation through solvent evaporation (so called drop-casting). A 3 cm x 3 cm crystalline-silicon (*c*-Si) substrate, which was already washed with isopropyl alcohol (IPA), acetone, ethanol, and DI water in successive way, was coated with polydimethylsiloxane (PDMS) using spin coating (9000 rpm for 60 s) and thermally cured in 80°C oven for 1 h. Then, 1 mL of a 10 wt% silica colloidal solution was drop-casted on the PDMS/*c*-Si substrate in a 24°C oven. During the evaporation of ethanol, silica nanospheres were self-assembled into opaline structures (see **Fig. S2**).

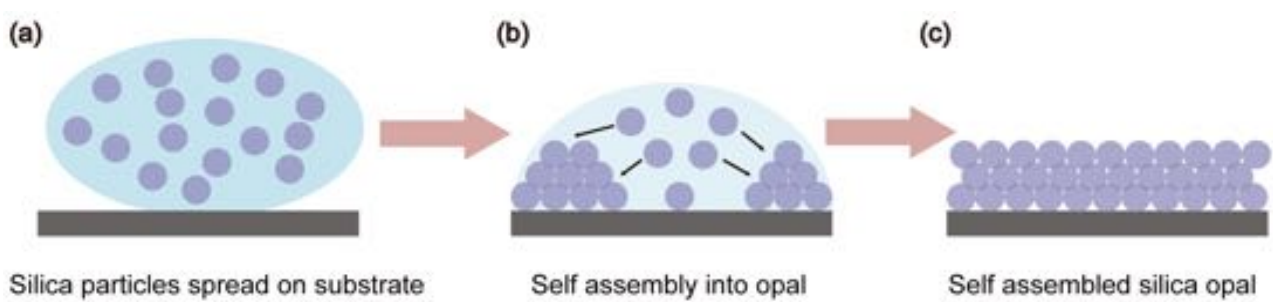

**Figure S2.** (a) Drop-casting of silica colloidal solution on the substrate. (b) By gravity, silica nanospheres are self-assembled on the substrate. (c) After the few hours, the solvent was completely evaporated and the assembled silica opals were obtained.

## 4. Microscopic view of the cracks in self-assembled opals

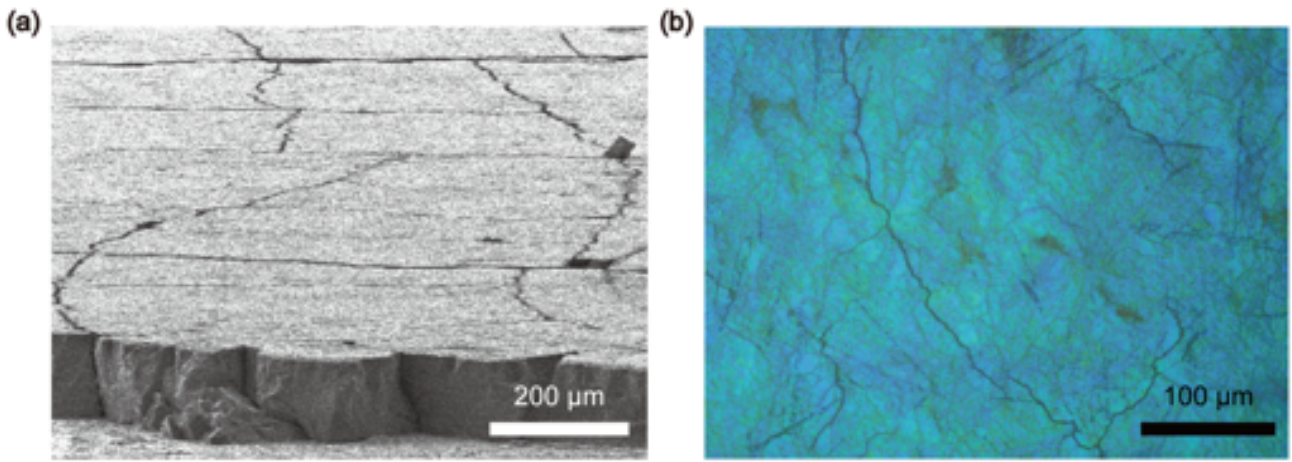

**Figure S3.** (a) Scanning electron microscope (SEM) and (b) bright field optical microscope (OM) images of the cracks in the self-assembled opals. The intrinsic cracks and point defects are clearly visible in both images, causing an unexpected diffusions and localizations of the incoming sunlight within assembled opals.

5. Experiments on radiative cooling

The measurement of outdoor radiative cooling was carried out in Seongbuk-gu, Seoul with 37°35'24.0"N 127°01'36.4"E (Korea University campus). Resistance temperature detectors (SA-1 RTD, Omega Korea) were attached to the backside of the samples to measure the temperature and connected to a data logger (OM-CP-OCTRCT, Omega Korea) to record the time-traced temperature variations. The RTD-attached samples were put into the custom-built, thermally isolated box (**Fig. S4**), consisting of paper box, aluminized polyethylene terephthalate (PET) film, styrofoam, acrylic box, and low-density polyethylene (LDPE) film (see **Figure 4b** of main manuscript). As a windshield, the LDPE seals the window. A pyranometer (TES1333R) was placed towards the sky to measure the solar irradiance incident on the sample.

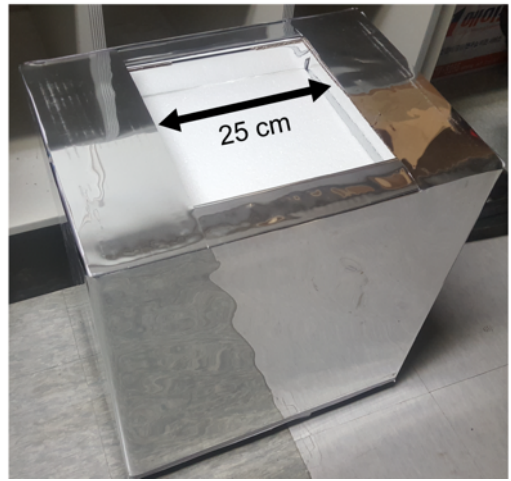

**Figure. S4** Outdoor temperature measurement set up without samples.

6. Calculation of radiative cooling power

We calculated the net cooling power as follows:[4]

$$P_{net}(T_{sample}) = P_{rad}(T_{sample}) - P_{atm}(T_{ambient}) - P_{sun} - P_{conduction+convection}, \quad (S1)$$

where each term in S1 indicate the total power of the sample, the power of the sample to radiation to the sky, the power absorbed by the sample from the atmosphere, the power absorbed by the sample from the sun, and the power due to the heat exchange with the surrounding environment (from left to right). Each term of the right side is detailed as follows.

$$P_{rad}(T_{sample}) = \int_0^{2\pi}\int_0^{\pi/2}\sin(\theta)\cos(\theta)\,d\theta d\phi \int_0^{\infty} I_{BB}(T_{sample},\lambda)\epsilon(\lambda,\theta)d\lambda, \quad (S2)$$

$$P_{atm}(T_{ambient}) = \int_0^{2\pi}\int_0^{\pi/2}\sin(\theta)\cos(\theta)\,d\theta d\phi \int_0^{\infty} I_{BB}(T_{ambient},\lambda)\epsilon_{atm}(\lambda,\theta)\alpha(\lambda,\theta)d\lambda, \quad (S3)$$

$$P_{sun} = \int_0^{\infty} I_{AM1.5}(\lambda)\alpha(\lambda,\theta)d\lambda, \quad (S4)$$

$$P_{conduction+convection} = h_c(T_{sample} - T_{ambient}), \quad (S5)$$

$\int_0^{2\pi}\int_0^{\pi/2}\sin(\theta)\cos(\theta)\,d\theta d\phi$ is a solid angle within which the sample absorbs or emits heat. $\epsilon$, $\alpha$, $h$, $k_B$, $c$, $\lambda$, and $I_{BB}(T,\lambda)$ indicate respectively emissivity, absorptivity, Planck's constant, Boltzmann constant, speed of light, the wavelength of the radiation wave, and blackbody radiation (i.e., the amount of emitted energy per unit wavelength at different wavelengths and temperature). Note that absorptivity should be equal to emissivity according to Kirchhoff's law. The emissivity in the equation **S3** was calculated as $\epsilon_{atm}(\lambda,\theta) = 1 - t(\lambda)^{1/\cos(\theta)}$, where $t(\lambda)$ is the terrestrial transparency window in the zenith direction. In equation **S4**, $I_{AM1.5}(\lambda)$ is the AM1.5 spectrum of the sunlight. Solar spectrum and terrestrial transparency window appeared in **Fig. 1d** of main manuscript.

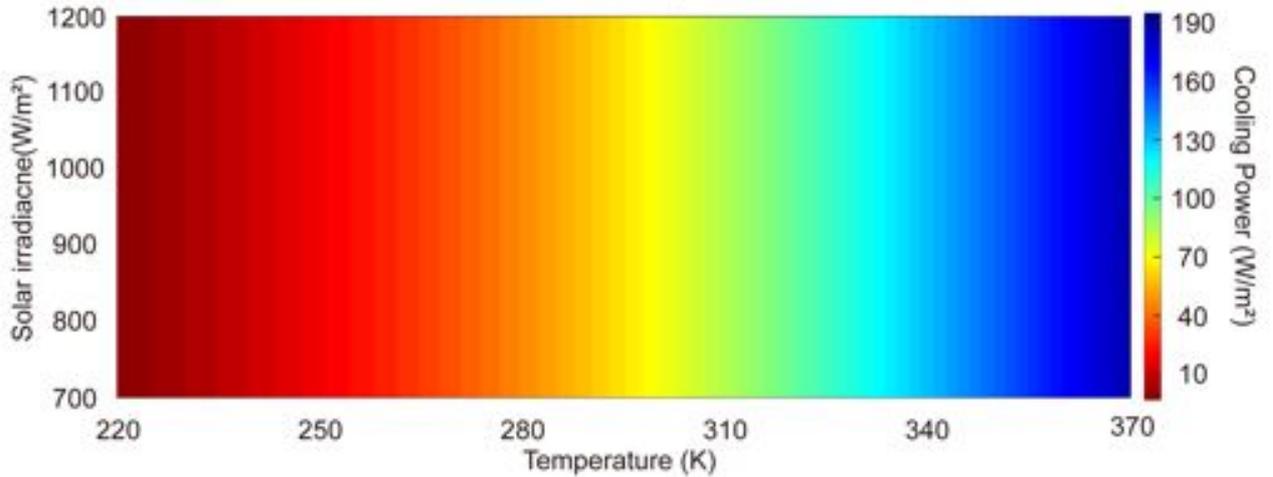

**Figure S5.** The calculated net cooling power of reddish opals according to the device temperature and solar irradiance.

7. Additional data for daytime radiative cooling of opals

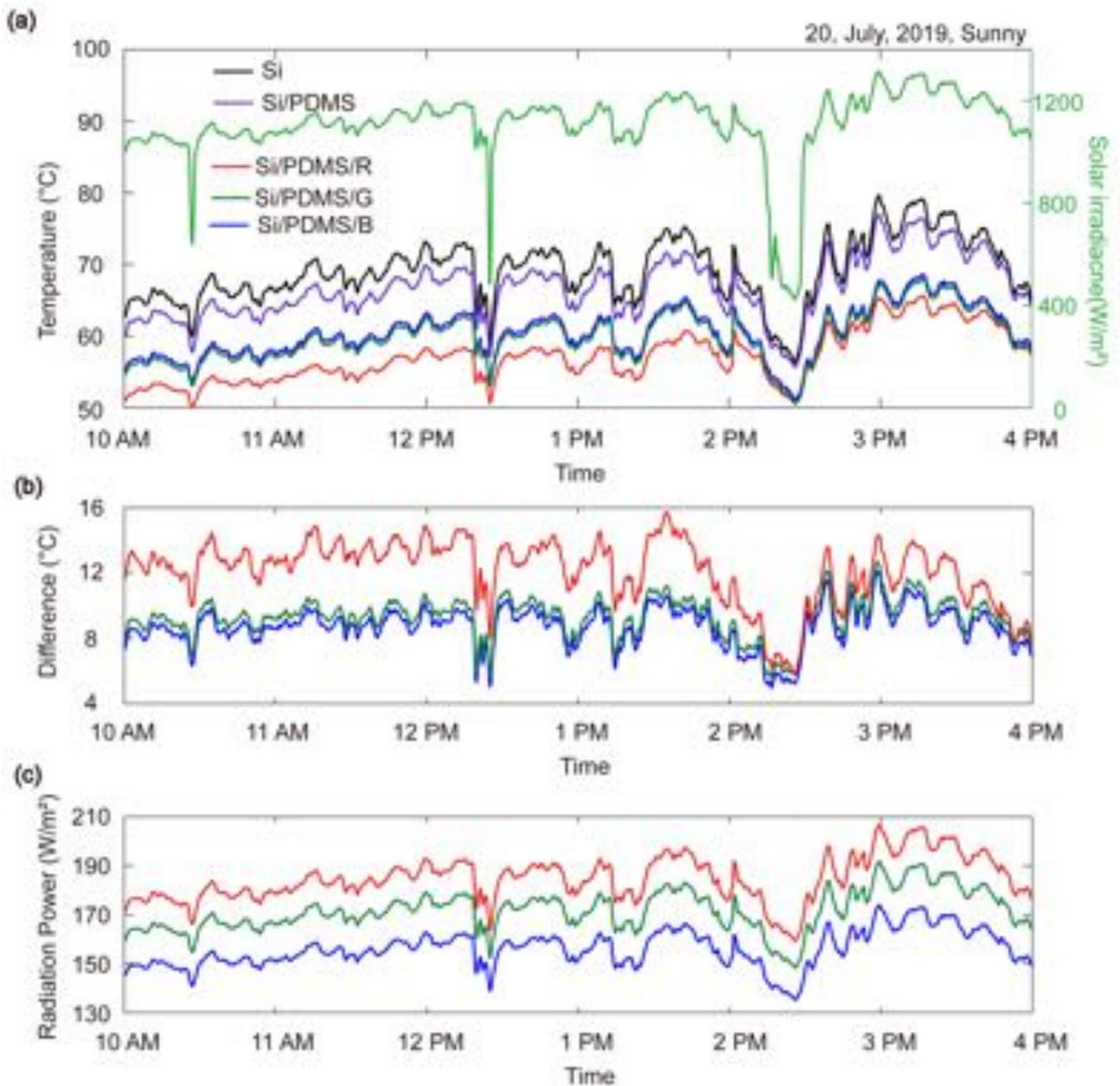

**Figure S6.** Additional time-traced temperature measurement for outdoor daytime radiative cooling experiment. These experiments were performed during sunny daytime on July 20[th], 2019. In that day, it was a quite sunny; but, cloud was temporarily passed over the sky. (a) Time-trace temperature variation. (b) Time-trance temperature difference between bare substrate and opals/substrate stacks. (c) Corresponding radiation power. (d) Radiation power of RGB opals as function of the device temperature and thermal coefficients.

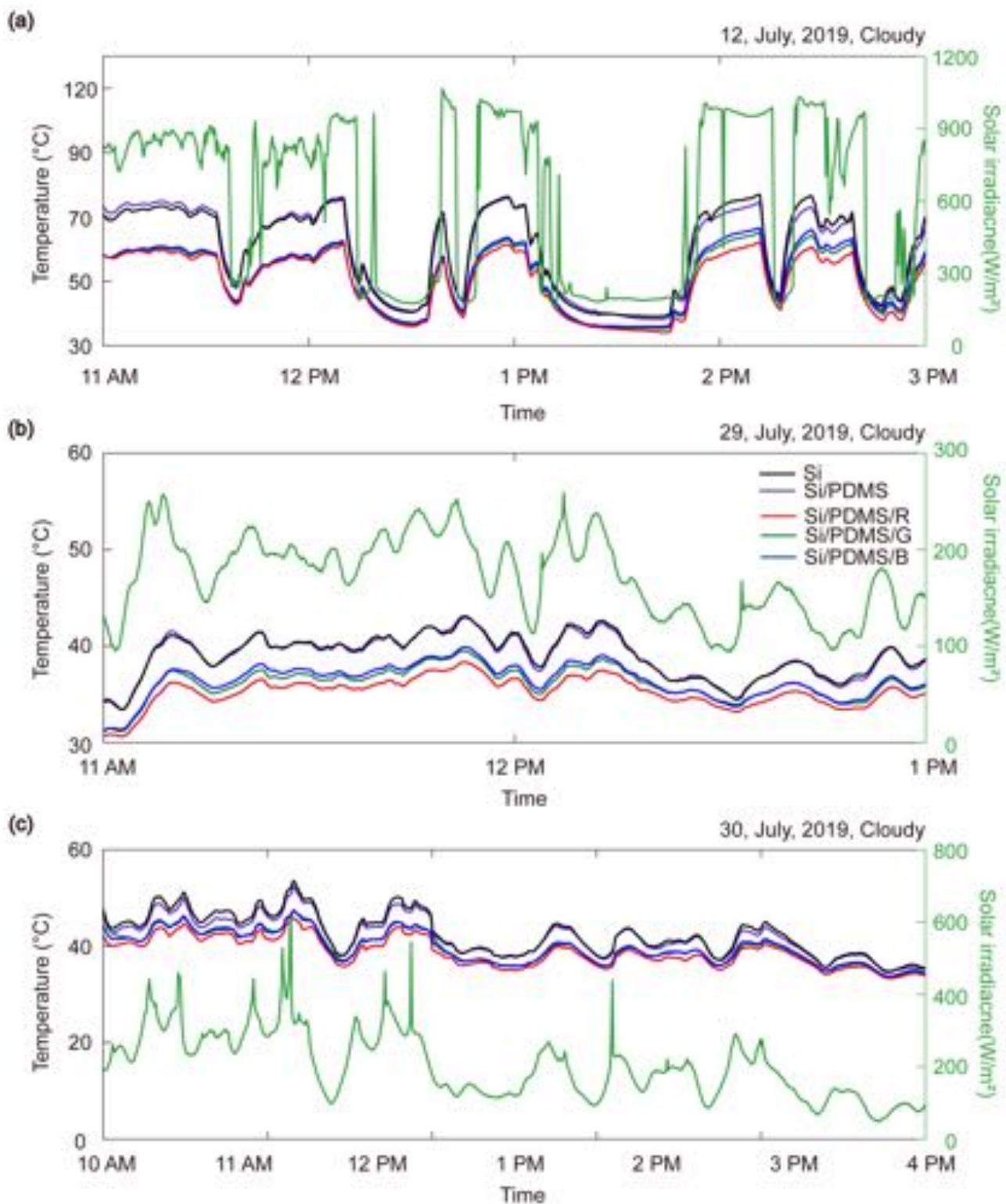

**Figure S7.** Additional time-traced temperature measurement for outdoor daytime radiative cooling experiment. These experiments were performed during cloudy daytime on (a) July 12[th], (b) 29[th], and (c) 30[th], 2019.